\documentclass[prb,aps,twocolumn,showpacs,nofootinbib]{revtex4}
\usepackage{graphicx} 
\usepackage{dcolumn} 
\usepackage{bm} 
\usepackage{color}

\begin{document}

\title{ARPES Spectral Function in Lightly Doped and Antiferromagnetically Ordered YBa$_{2}$Cu$_{3}$O$_{6+y}$}

\author{Wei Chen$^{1}$, Oleg P. Sushkov$^{1}$, and Takami Tohyama$^{2}$}

\affiliation{$^{1}$School of Physics, University of New South Wales, Sydney 2052, Australia \\
$^{2}$Yukawa Institute for Theoretical Physics, Kyoto University, Kyoto 606-8502, Japan}

\date{\today}

\begin{abstract}
At doping below $6\%$ the bilayer cuprate YBa$_{2}$Cu$_{3}$O$_{6+y}$
is a collinear antiferromagnet. Independent of doping the value of the staggered 
magnetization at  zero temperature is about $0.6\mu_B$.
 This is the maximum value  of the magnetization allowed by quantum fluctuations of 
localized spins.  In this low doping regime the compound is 
a normal  conductor with a finite  resistivity at zero temperature. 
These experimental observations create a unique opportunity for theory to perform a 
controlled calculation of the electron spectral function.
 In the present work we perform this calculation
within the framework of the extended $t-J$ model. 
As one expects the Fermi surface consists of small hole pockets centered  at
$(\pm\pi/2,\pm\pi/2)$. The electron spectral function is very strongly anisotropic
with maximum of intensity located at the inner parts of the pockets and with very small
intensity at the outer parts.
We also found that the antiferromagnetic correlations act against the bilayer 
bonding-antibonding splitting destroying it. The bilayer Fermi surface 
splitting is practically  zero. 
\end{abstract}


\maketitle

\section{Introduction}

One of the central issues in the physics of cuprates is the evolution of 
Fermi surface upon doping. Despite a consensus on the large Fermi surface 
in the overdoped side, the shape and nature of the Fermi surface
 in underdoped cuprates remains an unresolved issue.
Cuprates are doped Mott insulators. Theoretically there is
no doubt that minima of the dispersion of a single hole injected in
the Mott insulator are at the nodal points $(\pm\pi/2,\pm\pi/2)$.
This conclusion is supported by the Angle-Resolved Photoemission 
Spectroscopy (ARPES) data from undoped 
cuprates.~\cite{Damascelli03,Yoshida07}
This implies that at a sufficiently small doping holes
must go to the hole pockets. This is the small Fermi surface situation.
In real compounds disorder complicates the situation and significantly
masks the generic physics.
For example the prototypical cuprate La$_{2-x}$Sr$_x$CuO$_4$ is so disordered
that holes are strongly localized at doping $x\lesssim 0.1$.
In  the case of strong localization the notion of the Fermi surface is 
ambiguous. ARPES data from underdoped La$_{2-x}$Sr$_x$CuO$_4$ reveal  
Fermi arcs~\cite{Damascelli03,Yoshida07}
that are not consistent either with the small (hole pockets)
nor with the large Fermi surface.
On the other hand the very recent ARPES data from underdoped 
Bi$_2$Sr$_2$CaCu$_2$O$_{8+\delta}$ indicate small hole pockets.~\cite{Yang}

YBa$_{2}$Cu$_{3}$O$_{6+y}$ (YBCO) is probably the least disordered cuprate
in the low doping regime. In this paper we denote doping $x$ to be the hole 
concentration per unit cell of the CuO$_2$ layer.
YBCO is not superconducting below $x\lesssim 0.06$, where the compound 
remains a normal conductor with delocalized holes.
 The zero temperature resistivity remains finite,~\cite{WO01}
apart of a very weak logarithmic temperature dependence~\cite{sun,doiron}
expected for a weak disorder. 
The heat conductivity also indicates delocalization of 
holes.~\cite{sutherland05}
We emphasize that this is very different from La$_{2-x}$Sr$_x$CuO$_4$ where holes
are localized and hence the compound is the
Anderson insulator~\cite{Boebinger96,ando02} at
$x \lesssim 0.1$.~\cite{com1}
Ultimately, at the very low doping, $x\lesssim 0.01$, the disorder
wins even in YBCO and it also becomes the Anderson insulator.~\cite{WO01}
It is helpful to have in mind an approximate empiric formula~\cite{WO01,Liang}
\begin{equation}
\label{xy}
x\approx 0.35(y-6.20)
\end{equation}
to relate the doping level $x$ and the oxygen content $y$ in 
underdoped YBa$_{2}$Cu$_{3}$O$_{6+y}$ at $x \lesssim 0.12$.
The homogeneity of YBCO is the reason why Magnetic Quantum Oscillations (MQO) 
were observed in this compound.~\cite{Doiron-Leyraud07,LeBoeuf07,
Yelland08,Jaudet08,Sebastian08,Audouard09,Sebastian09,Sebastian09_2,Singleton09} 
The oscillations clearly indicate small Fermi pockets, while strictly speaking
nature of the pockets, including the sign of the charge of the fermion, experimentally
remains a controversial issue.

The collinear antiferromagnetism (AF) in YBCO is preserved  up to
the doping level $x\approx 0.06$. Moreover, the zero temperature staggered magnetization $\mu\approx 0.6\mu_B$
is practically doping-independent, having the same value as in the parent Mott insulator.~\cite{con10} This is the maximum value of magnetization allowed by quantum fluctuations of 
localized spins. 
The doping behavior of the staggered magnetization in YBCO
is very different from that in La$_{2-x}$Sr$_x$CuO$_4$ where
the staggered magnetization decays dramatically with doping. 
A special mechanism has been proposed~\cite{sushkov11} to explain antiferromagnetism 
in YBCO at $x \lesssim 0.06$.
For purposes of the present work details of the mechanism are not
important. The only important point is that the antiferromagnetism is
independent of doping. This direct experimental observation in
combination with simple metallic behaviour (finite resistivity at zero
temperature that is another direct experimental observation)
gives a unique opportunity to perform a controlled and fully reliable 
theoretical calculation of the electron spectral function at $x \lesssim 0.06$.
In the present work we perform this calculation.

Our analysis is based on the  $t-t'-t''-J$ model and employs the 
self-consistent Born approximation (SCBA). 
SCBA has been widely applied to study a single hole dressed by spin 
fluctuations.~\cite{Sushkov97,Kane89,Martinez91,Liu92,Ramsak92,nazarenko} 
On the other hand  application of the method at finite doping has been 
very limited~\cite{Igarashi92,Kyung96} because a usual
justification of the method requires a long range AF order,
and the common wisdom is that even a tiny doping destroys the  order.
AF ordered YBCO does not comply with the common wisdom and
provides a unique opportunity to address spin fluctuations 
up to very high accuracy. The AF order implies that the single loop vertex 
correction  is forbidden (the "Migdal theorem''), so SCBA is exact up to 
double loop corrections.~\cite{Liu92}

Another important property of the  quasiparticle dispersion is the layer bonding-antibonding 
splitting in the bilayer cuprate. A density functional theory calculation shows a more or less constant splitting through out the whole Brillouin zone.~\cite{Andersen95}
ARPES in overdoped YBCO confirms the splitting.~\cite{borisenko06,Fournier10}
On the other hand, in the underdoped regime 
ARPES measurements~\cite{Fournier10} indicate no such splitting.
To explain this in the present paper we show that antiferromagnetic correlations between layers 
in the bilayer system diminish the splitting in spite of the strong chemical tendency towards 
the splitting. The similar result for undoped YBCO was previously obtained in 
Ref.~\onlinecite{nazarenko}.
The antiferromagnetic correlations are due to the antiferromegnetic exchange $J_{\perp}\approx 10meV$
between the layers.~\cite{Reznik96} 
It is worth noting that the small hole pockets are essential for the
suppression of the bilayer splitting in the underdoped regime.

In the present work we consider the clean limit without any disorder.
There are recent ARPES experiments on underdoped YBCO~\cite{Hossain08,Fournier10} where the mechanism of doping is related to depositing of potassium atoms on the surface.
The potassium ions give rise to a random potential for mobile holes. 
The degree of disorder in these experiments remains an open issue when compared with the present results.

The structure of the paper is the following. In Section II, we introduce the Hamiltonian
and calculate vertexes for many-body diagrammatic technique.
In Section III the method of SCBA at finite doping is discussed. The method
implies the spin-charge separation.
The spin-charge recombination amplitude and the ARPES spectral function is calculated in Section IV.
Section V summarizes our results.

\section{The extended $t-J$ model for the bilayer YBCO}

\begin{figure}[ht]
\vspace{-5pt}
\begin{center}
\includegraphics[clip=true,width=0.8\columnwidth]{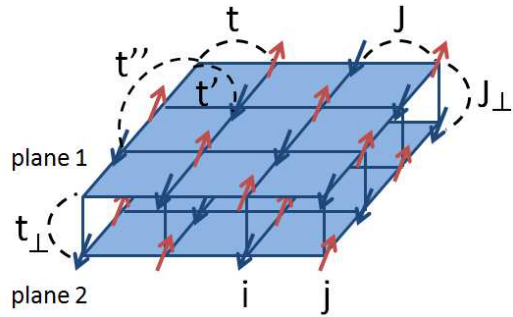}
\caption{(color online) Schematics of double layer Hamiltonian, Eq.~(\ref{double_layer_H}), and definition of coordinate, Eq.~(\ref{ij_definition}).}
\label{fig:double_layer_structure}
\end{center}
\end{figure}

We simulate lightly doped YBCO by using the double layer $t-t^{\prime}-t^{\prime\prime}-J$ model 
with constant interlayer coupling $t_{\perp}$ and $J_{\perp}$
\begin{eqnarray}
H&=&-t\sum_{m,\langle ij\rangle,\sigma}c_{i,m,\sigma}^{\dag}c_{j,m,\sigma}
-t^{\prime}\sum_{m,\langle ij\rangle^{\prime},\sigma}c_{i,m,\sigma}^{\dag}c_{j,m,\sigma}
\nonumber \\
&&-t^{\prime\prime}\sum_{m,\langle ij\rangle^{\prime\prime},\sigma}c_{i,m,\sigma}^{\dag}c_{j,m,\sigma}+J\sum_{\langle ij\rangle}{\bf S}_{i,m}\cdot{\bf S}_{j,m}
\nonumber \\
&&-t_{\perp}\sum_{l,\sigma}\left(c_{l,1,\sigma}^{\dag}c_{l,2,\sigma}+h.c.\right)
+J_{\perp}\sum_{l}{\bf S}_{l,1}\cdot{\bf S}_{l,2}
\nonumber \\
&=&H_{t^{\prime},t^{\prime\prime}}+H_{J,J_{\perp}}+H_{t,t_{\perp}}
\label{double_layer_H}
\end{eqnarray}
where $t/t^{\prime}/t^{\prime\prime}$ is the nearest/next-nearest/next-next-nearest neighbor in-plane hopping, respectively, and $m=\left\{1,2\right\}$ is the plane index. Since $J_{\perp}\approx 10meV >0$, the magnetic ordering along $c-$axis is also AF. We denote the projected coordinate $\left\{i,j\right\}$ in each plane as
\begin{eqnarray}
&&i\;\in\;\uparrow{\rm sublattice \;plane\;1} \ \ \ \equiv \ \ \ 
\downarrow{\rm sublattice \;plane\;2}\;
\nonumber \\
&&j\;\in\;\downarrow{\rm sublattice \;plane\;1}\ \ \ \equiv \ \ \ 
\uparrow{\rm sublattice \;plane\;2}\;
\nonumber \\
&&l\in\left\{i,j\right\}
\label{ij_definition}
\end{eqnarray}
 Through out the article we set energy unit as $J=130$meV$\rightarrow 1$,
hence $J_{\perp} \approx 0.08$. Schematics of the model is shown in 
Fig.~\ref{fig:double_layer_structure}.

We use the following values of the intralayer hopping parameters
$t=3.1$, $t^{\prime}=-0.5$, $t^{\prime\prime}=0.4$.
While the value of $t$ corresponds to that obtained in
the density functional theory calculation,~\cite{Andersen95}
values of $t^{\prime}=-0.5$ and $t^{\prime\prime}=0.4$ are
somewhat different.
Ref.~\onlinecite{Andersen95} gives for the optimally doped
YBCO the following values, $t^{\prime}\approx 0.8$, $t^{\prime\prime}\approx -0.7$.
Based on our results that we compare with ARPES data
we believe that the values of $t^{\prime}=-0.5$ and $t^{\prime\prime}=0.4$
accepted in the present work are more suitable for underdoped YBCO.
However, in the end the difference between these two sets of parameters
is not qualitatively important.
We take a constant interlayer tunneling $t_{\perp}$ based again on 
the first principle calculation,~\cite{Andersen95}
which shows a practically constant splitting between bonding and 
antibonding bands with the value of the splitting
corresponding to $t_{\perp} \approx 100meV \to 1$.
ARPES data from overdoped YBCO~\cite{borisenko06,Fournier10}
support this splitting.
The estimate $t_{\perp}\approx 1$ is supported also by the ratio
of the superexchange parameters, $0.08 =J_{\perp}/J=t_{\perp}^{2}/t^{2}$.

Following the standard SCBA philosophy the Hamiltonian (\ref{double_layer_H})
is grouped  into three sectors, $H_{t^{\prime},t^{\prime\prime}}$, $H_{J,J_{\perp}}$, and $H_{t,t_{\perp}}$, 
which correspond to bare hole, bare magnon, and hole-magnon interaction, respectively.
We first discuss the bare magnon sector $H_{J,J_{\perp}}$. The Holstein-Primakoff bosons are 
defined according to sublattices in each layer, as defined in Eq.~(\ref{ij_definition}). 
In plane $1$
\begin{eqnarray}
&&S_{i,1}^{z}=\frac{1}{2}-a_{i,1}^{\dag}a_{i,1}
\nonumber \\
&&S_{i,1}^{+}=a_{i,1}
\nonumber \\
&&S_{i,1}^{-}=a_{i,1}^{\dag}
\nonumber \\
&&S_{j,1}^{z}=-\frac{1}{2}+b_{j,1}^{\dag}b_{j,1}
\nonumber \\
&&S_{j,1}^{+}=b_{j,1}^{\dag}
\nonumber \\
&&S_{j,1}^{-}=b_{j,1}
\end{eqnarray}
and in plane 2
\begin{eqnarray}
&&S_{i,2}^{z}=-\frac{1}{2}+b_{i,2}^{\dag}b_{i,2}
\nonumber \\
&&S_{i,2}^{+}=b_{i,2}^{\dag}
\nonumber \\
&&S_{i,2}^{-}=b_{i,2}
\nonumber \\
&&S_{j,2}^{z}=\frac{1}{2}-a_{j,2}^{\dag}a_{j,2}
\nonumber \\
&&S_{j,2}^{+}=a_{j,2}
\nonumber \\
&&S_{j,2}^{-}=a_{j,2}^{\dag}\ .
\end{eqnarray}
The Fourier transform is
\begin{eqnarray}
&&a_{i,1}=\sqrt{\frac{2}{N}}\sum_{\bf q}a_{{\bf q},1}e^{i{\bf q\cdot r}_{i}}
\nonumber \\
&&b_{j,1}=\sqrt{\frac{2}{N}}\sum_{\bf q}b_{{\bf q},1}e^{i{\bf q\cdot r}_{j}}
\nonumber \\
&&a_{j,2}=\sqrt{\frac{2}{N}}\sum_{\bf q}a_{{\bf q},2}e^{i{\bf q\cdot r}_{j}}
\nonumber \\
&&b_{i,2}=\sqrt{\frac{2}{N}}\sum_{\bf q}b_{{\bf q},2}e^{i{\bf q\cdot r}_{i}} \ ,
\end{eqnarray}
where  summation over ${\bm q}$ is restricted inside 
Magnetic Brillouin Zone (MBZ).
By introducing parity $+$ and $-$ bases with respect to interchange of two planes
\begin{eqnarray}
&&a_{{\bf q},\pm}=\frac{1}{\sqrt{2}}\left(a_{{\bf q},1}\pm a_{{\bf q},2}\right)
\nonumber \\
&&b_{{\bf q},\pm}=\frac{1}{\sqrt{2}}\left(b_{{\bf q},1}\pm b_{{\bf q},2}\right)
\label{magnon_parity_basis}
\end{eqnarray}
and Bogoliubov transformation
\begin{eqnarray}
&&a_{{\bf q},\pm}=u_{{\bf q},\pm}\alpha_{{\bf q},\pm}+v_{{\bf q},\pm}\beta_{{\bf -q},\pm}^{\dag}
\nonumber \\
&&b_{{\bf -q},\pm}=v_{{\bf q},\pm}\alpha_{{\bf q},\pm}^{\dag}+u_{{\bf q},\pm}\beta_{{\bf -q},\pm}
\end{eqnarray}
one can diagonalize the Hamiltonian\cite{Tranquada89,Reznik96}
\begin{eqnarray}
H_{J,J_{\perp}}&=&\sum_{\bf q}\left(\alpha_{{\bf q},+}^{\dag}\alpha_{{\bf q},+}+\beta_{{\bf q},+}^{\dag}\beta_{{\bf q},+}\right)\omega_{{\bf q},+}
\nonumber \\
&+&\sum_{\bf q}\left(\alpha_{{\bf q},-}^{\dag}\alpha_{{\bf q},-}+\beta_{{\bf q},-}^{\dag}\beta_{{\bf q},-}\right)\omega_{{\bf q},-}\;,
\nonumber \\
\omega_{{\bf q},\pm}&=&2J\left\{(1+\frac{\alpha_{\perp}}{4})^{2}-(\gamma_{\bf q}\pm\frac{\alpha_{\perp}}{4})^{2}\right\}^{1/2}\nonumber\\
\gamma_{\bf q}&=&\frac{1}{2}\left(\cos q_x+\cos q_y\right) \ ,
\label{double_layer_dispersion}
\end{eqnarray}
where we denote $\alpha_{\perp}=J_{\perp}/J$. Here $\omega_{\bf q,+}$ mode is gapless at $(0,0)$ and gapped at $(\pi,\pi)$, while $\omega_{\bf q,-}$ mode is the opposite. This is consistent with the fact that optical mode is frequently referred to even parity, and acoustic mode to odd parity in inelastic neutron scattering
experiments which practically measure magnon dispersion near $(\pi,\pi)$.\cite{Tranquada89,Reznik96} The Bogoliubov coefficients are
\begin{eqnarray}
&&u_{{\bf q},\pm}=\sqrt{\frac{J+\frac{J_{\perp}}{4}}{\omega_{{\bf q},\pm}}+\frac{1}{2}}
\nonumber \\
&&v_{{\bf q},\pm}=-{\rm sign}({\gamma_{\bf q}\pm\frac{\alpha_{\perp}}{4}})\sqrt{\frac{J+\frac{J_{\perp}}{4}}{\omega_{{\bf q},\pm}}-\frac{1}{2}}
\end{eqnarray}
The magnon Green's function is defined as 
\begin{eqnarray}
D_{\pm}(\omega,{\bf q})&=&-i\int_{0}^{\infty}\langle T\alpha_{\bf q,\pm}(t)\alpha_{\bf q,\pm}^{\dag}(0)\rangle e^{i\omega t}dt
\nonumber \\
&=&-i\int_{0}^{\infty}\langle T\beta_{\bf q,\pm}(t)\beta_{\bf q,\pm}^{\dag}(0)\rangle e^{i\omega t}dt
\nonumber \\
&=&\frac{1}{\omega-\omega_{\bf q,\pm}+i\eta}\;.
\end{eqnarray}
In the present work we do not consider renormalization of magnon dispersion 
due to interaction 
with holes, but simply adopt the magnon sector of undoped Mott insulator.
This assumption is justified because $\mu$SR measurements indicate almost 
unrenormalized staggered magnetization up to doping level 
$x\approx 0.06$.\cite{con10}
The magnon dispersion close to the Goldstone point
${\bf q}=0$ [equivalent to ${\bf q}= (\pm\pi,\pm\pi)$] is somewhat 
changed under  doping.~\cite{sushkov11}
However, due to the Adler's theorem, magnons close to the Goldstone point 
practically do not influence the hole dispersion, see discussion below.
The main contribution to the hole dispersion comes from magnons that are 
far away from the Goldstone point. For this regime, resonant inelastic 
X-ray scattering(RIXS) clearly demonstrates 
that magnons are almost independent of doping.~\cite{RIX}
Theoretical analysis of the RIXS experimental data is an important 
problem that will be considered elsewhere.~\cite{wchen11} 

\begin{figure}[ht]
\vspace{-5pt}
\begin{center}
\includegraphics[clip=true,width=0.95\columnwidth]{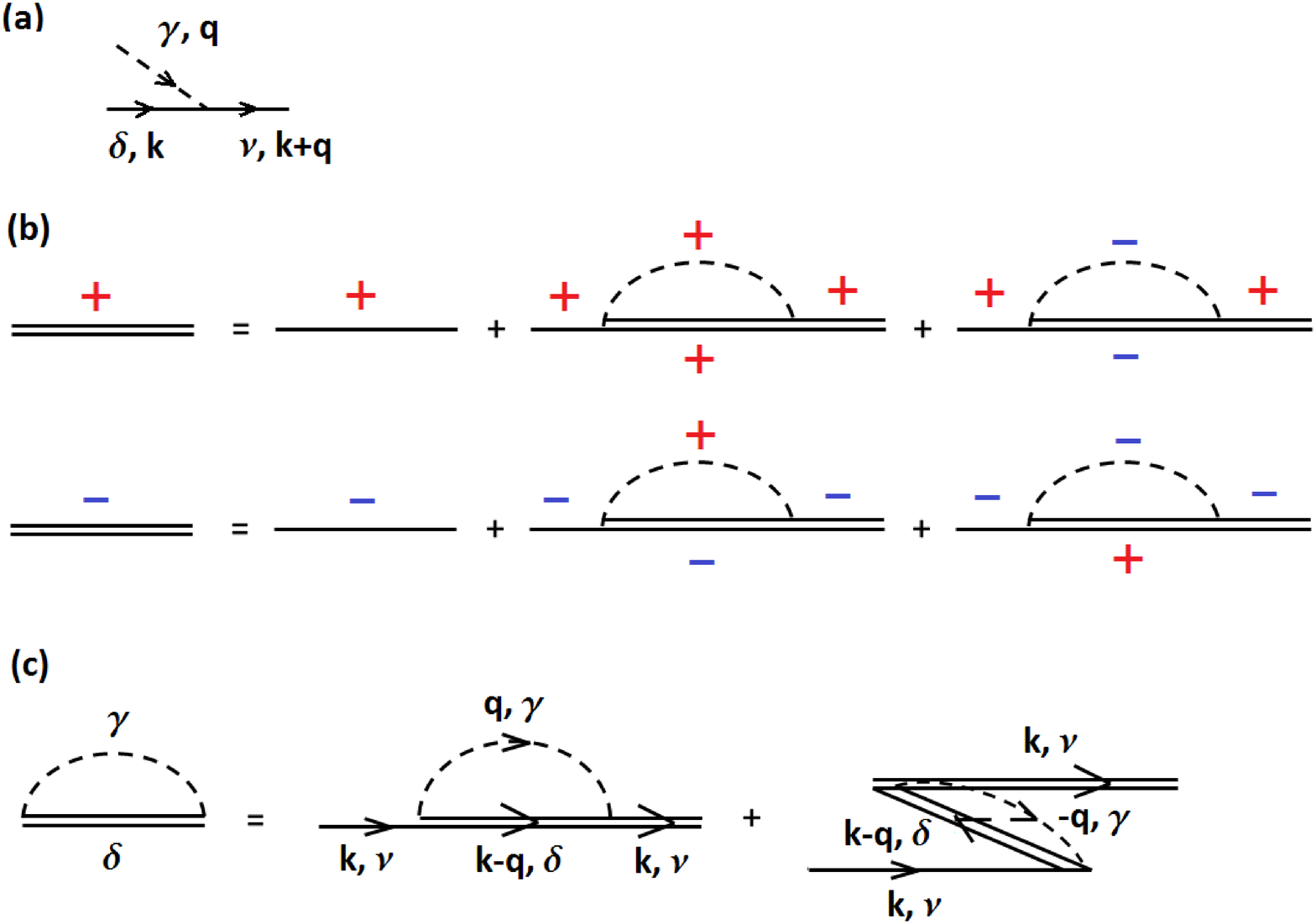}
\caption{(color online) Feynman diagrams for (a) the hole-magnon vertex defined in 
Eq.~(\ref{hole_magnon_vertices}), (b) Dyson's equation Eq.~(\ref{Gd_SCBA}) for $G_{d\pm}$, and (c) schematics of advanced and retarded part of each 
self-energy diagram, as calculated in Eq.~(\ref{hole_magnon_S}). }
\label{fig:Gd_Feynman_diagrams}
\end{center}
\vspace{-25pt}
\end{figure}

To address the bare hole dispersion from the $H_{t^{\prime},t^{\prime\prime}}$ term, we first define hole operators according to the coordinate in Eq.~(\ref{ij_definition})
\begin{eqnarray}
&&d_{{\bf k},1,\uparrow}=\sqrt{\frac{2}{N(1/2+m)}}\sum_{j}c_{j,1,\downarrow}^{\dag}e^{-i{\bf k\cdot r}_{j}}
\nonumber \\
&&d_{{\bf k},1,\downarrow}=\sqrt{\frac{2}{N(1/2+m)}}\sum_{i}c_{i,1,\uparrow}^{\dag}e^{-i{\bf k\cdot r}_{i}}
\nonumber \\
&&d_{{\bf k},2,\uparrow}=\sqrt{\frac{2}{N(1/2+m)}}\sum_{i}c_{i,2,\downarrow}^{\dag}e^{-i{\bf k\cdot r}_{i}}
\nonumber \\
&&d_{{\bf k},2,\downarrow}=\sqrt{\frac{2}{N(1/2+m)}}\sum_{j}c_{j,2,\uparrow}^{\dag}e^{-i{\bf k\cdot r}_{j}} \ ,
\end{eqnarray}
where $m=|\langle S_z\rangle|\approx 0.3$, see Ref.~\onlinecite{Sushkov97}.
The fixed parity states are
\begin{eqnarray}
d_{{\bf k},\pm,\sigma}=\frac{1}{\sqrt{2}}\left(d_{{\bf k},1,\sigma}\pm d_{{\bf k},2,\sigma}\right)
\label{hole_parity_basis}
\end{eqnarray}
The bare hole dispersion is then
\begin{eqnarray}
&&H_{t^{\prime},t^{\prime\prime}}=\sum_{{\bf k},\sigma}\epsilon_{{\bf k},+}^{0}d_{{\bf k},+,\sigma}^{\dag}d_{{\bf k},+,\sigma}+\sum_{{\bf k},\sigma}\epsilon_{{\bf k},-}^{0}d_{{\bf k},-,\sigma}^{\dag}d_{{\bf k},-,\sigma}
\nonumber \\
&&\epsilon_{{\bf k},\pm}^{0}=\epsilon_{{\bf k}}^{0}=4t^{\prime}\cos k_{x}\cos k_{y}+2t^{\prime\prime}\left(\cos 2k_{x}+\cos 2k_{y}\right)
\nonumber \\
\end{eqnarray}
Since site $i$ on plane $1$ and site $i$ on plane $2$ belong to different sublattices, the interlayer 
hopping does not enter the bare dispersion. This is the reason why the AF
correlations suppress the interlayer  hopping.
Similar to the in-plane nearest-neighbor hopping, the interlayer hopping contains the
spin-flip process that contributes to the hole-magnon interaction. The corresponding vertex 
therefore contains contributions from both $t$ and $t_{\perp}$, as calculated in Appendix A. 
The Hamiltonian $H_{t,t_{\perp}}$ and the corresponding vertex read
\begin{eqnarray}
&&H_{t,t_{\perp}}=\sum_{{\bf k,q},\gamma\delta\nu}g_{{\bf k,q},\gamma\delta}\left(d_{{\bf k+q},\nu,\downarrow}^{\dag}d_{{\bf k},\delta,\uparrow}\alpha_{{\bf q},\gamma}\right.
\nonumber \\
&&\;\;\;\;\;\;\;\;\;\;\;\left.+d_{{\bf k+q},\nu,\uparrow}^{\dag}d_{{\bf k},\delta,\downarrow}\beta_{{\bf q},\gamma}\right)+h.c.\;,
\nonumber \\
&&g_{{\bf k,q},\gamma\delta}=\frac{4t}{\sqrt{N}}\left(\gamma_{\bf k}u_{{\bf q},\gamma}+\gamma_{\bf k+q}v_{{\bf q},\gamma}\right)
\nonumber \\
&&\;\;\;\;\;\;\;+\delta\frac{t_{\perp}}{\sqrt{N}}\left(u_{{\bf q},\gamma}+\gamma v_{{\bf q},\gamma}\right)\;.
\label{hole_magnon_vertices}
\end{eqnarray}
We denote the parity index by $\gamma,\delta,\nu=\pm 1$.
This is the parity of the annihilated magnon, the annihilated hole, and the created hole, as shown 
in Fig.~\ref{fig:Gd_Feynman_diagrams}(a). Conservation of parity implies $\nu=\gamma\delta$. 
The vertex (\ref{hole_magnon_vertices}) is zero at $q=0$, this is a consequence of Adler's
theorem, and this is why the contribution of magnons with small momenta to the self-energy
is negligible.

\section{Self Consistent Born Approximation}
The SCBA is based on summation of diagrams having the highest possible
power of the hopping parameter $t$ at a given number of 
loops.~\cite{Sushkov97,Kane89,Martinez91,Liu92,Ramsak92,nazarenko} 
This implies that the method is justified at $t/J > 1$.
A very important point is that due to the spin structure of the theory
(conservation of $S_z$) there is no single loop correction to the hole-magnon 
vertex shown in Fig.~\ref{fig:Gd_Feynman_diagrams}(a).
Therefore, the  diagrams having the highest possible
power of  $t$ are the rainbow diagrams. Thus, SCBA is summation of rainbow
diagrams. Again, the method is justified due to (i) absence of the single loop
vertex correction, (ii) due to $t/J > 1$.

Compared to the single hole in a single layer case,~\cite{Sushkov97,Kane89,Martinez91,Liu92,Ramsak92} the
present calculation has two complications: the double layer and the finite doping.
Importantly, since we consider the collinear AF state,
the both points (i) and (ii) justifying the method are still valid
in spite of the complications.
We stress that although the finite doping case has been discussed before,~\cite{Igarashi92,Kyung96} it is not until current calculation that 
SCBA is rigorously formulated for a real material that AF order persists at finite doping.

At finite doping we have to use the  Feynman Green's function of the hole
\begin{eqnarray}
G_{d\sigma\pm}(\epsilon,{\bf k})=-i\int_{-\infty}^{\infty}\langle Td_{{\bf k},\pm,\sigma}(t)d_{{\bf k},\pm,\sigma}^{\dag}(0)\rangle e^{i\epsilon t}dt
\label{hole_GF}
\end{eqnarray}
instead of the retarded Green's function in the undoped 
case.~\cite{Sushkov97,Kane89,Martinez91,Liu92,Ramsak92,nazarenko}
The Green's function (\ref{hole_GF}) has the parity index $\pm$ reflecting the
double layer structure, and the pseudospin index $\sigma=\uparrow\downarrow$ 
reflecting the AF structure.
Since the up and down pseudospins are degenerate, we omit the psedospin index 
$\sigma$ for the rest of the article, although one should keep in mind that 
vertexes in Eq.~(\ref{hole_magnon_vertices}) always flip the pseudospin. 
In the calculation of self-energy, we adopt the spectral representation 
\begin{eqnarray}
G_{d\pm}(\epsilon,{\bf k})
&=&\int_{0}^{\infty}dx\frac{ A_{\pm}(x,{\bf k})}{\epsilon-x+i0}+\int_{-\infty}^{0}dx\frac{ B_{\pm}(x,{\bf k})}{\epsilon-x-i0}\;.
\nonumber \\
&&\
\label{Gd_spectral_representation}
\end{eqnarray}
The chemical potential is set equal to zero.
The technical advantage of using spectral representation is that it ensures the causality of 
Dyson's equation in any order, as we calculate the self-energy in terms of the 
spectral functions. This is important because Dyson's equation typically converges 
at about $10 \sim 20$th order, therefore causality must be ensured at each order. 
The doping is
\begin{eqnarray}
\label{doping}
2x=2\sum_{\bf k}\sum_{\pm}\int_{-\infty}^{0}dxB_{\pm}(x,{\bf k}) \ .
\end{eqnarray}
The summation over ${\bf k}$ is limited inside the MBZ.
The coefficient $2$ in the left-hand-side of (\ref{doping})
is due to the bilayer, the coefficient $2$
in the right-hand-side is due to pseudospin.

 Dyson's equations shown schematically  in Fig.~{\ref{fig:Gd_Feynman_diagrams}}(b) read
\begin{eqnarray}
&&G_{d+}(\epsilon,{\bf k})=\left[\epsilon-\epsilon_{\bf k,}^0-\Sigma_{++}(\epsilon,{\bf k})-\Sigma_{--}(\epsilon,{\bf k})+i0\right]^{-1}
\nonumber \\
&&G_{d-}(\epsilon,{\bf k})=\left[\epsilon-\epsilon_{\bf k,}^0-\Sigma_{+-}(\epsilon,{\bf k})-\Sigma_{-+}(\epsilon,{\bf k})+i0\right]^{-1}
\nonumber \\
&&
\label{Gd_SCBA}
\end{eqnarray}
The subscripts in the self-energy $\Sigma_{\pm,\pm}$ indicate parities of the intermediate hole
and magnon as it is shown   in Fig.~{\ref{fig:Gd_Feynman_diagrams}}(b).
Note that the considered theory does not possess the usual cross-leg-symmetry of 
vertexes. 
Therefore the usual Feynman technique is not sufficient,~\cite{Kyung96}
and one has to adopt the Goldstone-Brueckner technique that explicitly separates forward in time  and backward in time
diagrams, as shown in {\ref{fig:Gd_Feynman_diagrams}}(c).
Explicit expressions for the self-energy are derived in Appendix B
and presented in Eq.~(\ref{hole_magnon_S}).

We now discuss symmetry properties of Green's functions imposed by definitions of 
operators (\ref{magnon_parity_basis}) and (\ref{hole_parity_basis}). 
Due to the chequerboard AF order, hole and magnon operators change parity under translation by
the AF wave vector ${\bf Q}=(\pi,\pi)$, ${\bf k}\rightarrow {\bf k+Q}$.
\begin{eqnarray}
&&d_{{\bf k},+,\uparrow}=-d_{{\bf k+Q},-,\uparrow}
\nonumber \\
&&d_{{\bf k},+,\downarrow}=d_{{\bf k+Q},-,\downarrow}
\nonumber \\
&&\alpha_{{\bf q},+}=\alpha_{{\bf q},-}
\nonumber \\
&&\beta_{{\bf q},+}=-\beta_{{\bf q},-} \ .
\label{dispersion_degeneracy}
\end{eqnarray}
Hence the Green's functions satisfy the following symmetry relations
\begin{eqnarray}
&&G_{d+}(\epsilon,{\bf k+Q})=G_{d-}(\epsilon,{\bf k})\;,
\nonumber \\
&&D_{+}(\omega,{\bf q+Q})=D_{-}(\omega,{\bf q})\;.
\end{eqnarray}
As a result, the hole dispersion of parity $+$ and $-$ swap at the MBZ boundary,
as previously reported for the single hole case.~\cite{nazarenko}
 This implies 
that at the MBZ boundary, the magnon dispersion of either parity are degenerate, and the hole 
dispersion of either parity are also degenerate, as addressed below.

\begin{figure}[ht]
\begin{center}
\includegraphics[clip=true,width=0.9\columnwidth]{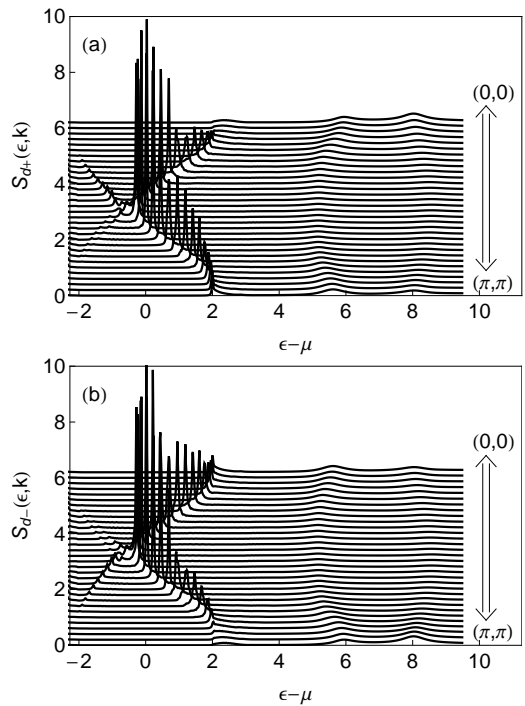}
\caption{Spectral functions $S_{d+}$ (a) and $S_{d-}$ (b) for different values
of ${\bf k}$ along the nodal direction from $(0,0)$ to $(\pi,\pi)$. 
There is an offset 0.2 between subsequent values of ${\bf k}$.
The doping level is $x=0.059$.
}
\label{fig:Gd_figure}
\end{center}
\end{figure}

\begin{figure}[ht]
\begin{center}
\includegraphics[clip=true,width=0.9\columnwidth]{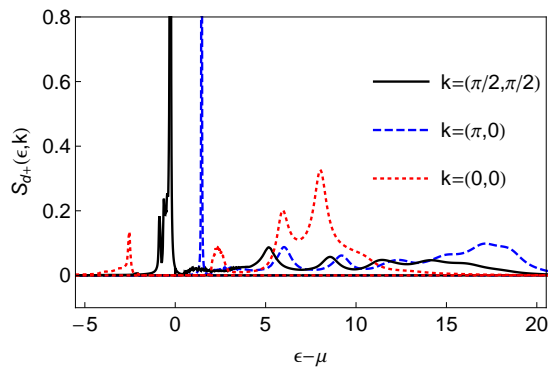}
\caption{Spectral functions $S_{d+}$ for 
${\bf k}=(0,0)$, ${\bf k}=(\pi/2,\pi/2)$ and ${\bf k}=(\pi,0)$.
The doping level is $x=0.059$.
 }
\label{fig:Gd_figure3p}
\end{center}
\end{figure}

Iterative numerical solution of Eqs.~(\ref{Gd_SCBA}),(\ref{hole_magnon_S}) 
requires more computational power compared to solution of similar equations
for undoped single layer case.~\cite{Sushkov97,Kane89,Martinez91,Liu92,Ramsak92,nazarenko}
Nevertheless the solution is  straightforward.
We solve Eq.~(\ref{Gd_SCBA}) inside MBZ on a $64\times 64$ cluster, with 
about 1700 frequency points (the frequency grid is $\Delta\omega=0.015$).
We present plots of the hole spectral function $S_d$ defined as $A(\epsilon,{\bm k})$
if $\epsilon >0$ and as $B(\epsilon,{\bm k})$ if $\epsilon < 0$, see 
Eq.~(\ref{Gd_spectral_representation}).
\begin{equation}
\label{sd}
S_{d\pm}(\epsilon,{\bm k})=A_{\pm}(\epsilon,{\bm k})+B_{\pm}(\epsilon,{\bm k}) \ .
\end{equation}
Figs.~\ref{fig:Gd_figure}(a) and (b) display the hole spectral functions 
(offset 0.2) of positive and negative parities for doping level $x=0.059$.
This doping is close to the highest doping level at which AF order persists while superconductivity 
has yet taken place.~\cite{con10} 
To show more details the spectral function $S_{d+}$ is plotted in 
Fig.~\ref {fig:Gd_figure3p}
for ${\bf k}=(0,0)$, ${\bf k}=(\pi/2,\pi/2)$ and ${\bf k}=(\pi,0)$. 
The spectral functions show a striking similarity to the hole spectral function in 
an undoped insulator: there is a well defined quasiparticle peak at low energy, and a 
large incoherent part  at higher energy.
In addition, a small but significant incoherent part is observed at energy below the 
quasiparticle peak, 
due to the electron plus multiple magnon configurations in the emission channel. 
Along nodal direction, the quasiparticle peak is most pronounced near $(\pi/2,\pi/2)$, 
and gradually decreases as approaching $(0,0)$ and $(\pi,\pi)$ where it is
dissolved in the hole-magnon continuum.

\begin{figure}[ht]
\begin{center}
\includegraphics[clip=true,width=0.9\columnwidth]{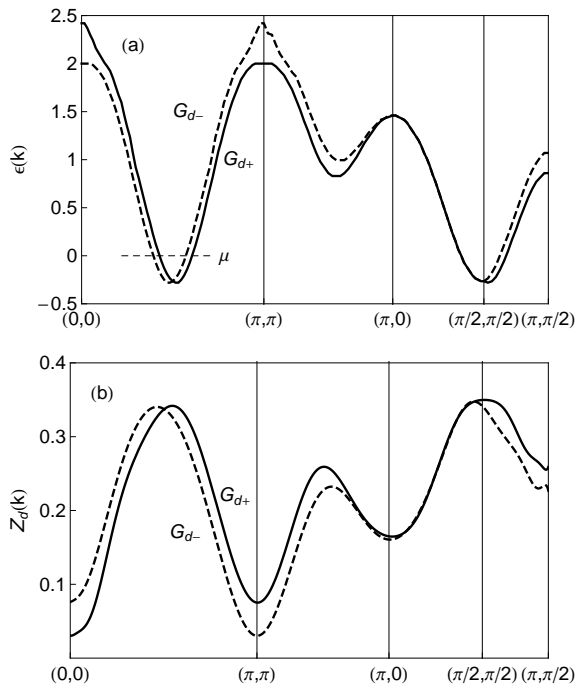}
\caption{
(a) The quasiparticle dispersion, and (b) the quasiparticle residue $Z_d$
along a certain line in BZ. The doping level is $x=0.059$. 
}
\label{fig:Gd_figure2}
\end{center}
\end{figure}

The position of the quasiparticle peak determines the quasiparticle dispersion $\epsilon_{\bf k}$.
The dispersion is shown in Fig.~\ref{fig:Gd_figure2}(a)  along a certain line 
in the Brillouin Zone (BZ).
Formation of small Fermi surface with hole pockets at ${\bf k}=(\pi/2,\pi/2)$ is
evident.
 Eq.~(\ref{dispersion_degeneracy}), imposes a nontrivial constraint on 
the dispersion: From ${\bf k}=(0,0)$ to $(\pi,\pi)$, one sees that the two parity bands are 
symmetric but swap at the MBZ boundary ${\bf k}=(\pi/2,\pi/2)$. Along the MBZ boundary, as shown 
in the section from ${\bf k}=(\pi,0)$ to $(\pi/2,\pi/2)$, 
the two bands are degenerate. At the Fermi surface $\epsilon_{{\bf k}+}$ differs from
$\epsilon_{{\bf k}-}$ by only $\approx 25$meV.
We remind that for the value $t_{\perp}=J=130$meV used in the calculation, one expects the splitting to be $260$meV. Therefore AF correlations strongly suppress the bilayer splitting.

In the section of $(\pi,\pi)$ to $(\pi,0)$ in Fig.~\ref{fig:Gd_figure2}(a), one sees a local 
dispersion minimum near $(\pi,\pi/2)$. However, in the $(\pi/2,\pi/2)$ to $(\pi,\pi/2)$ section, 
ones sees that $(\pi,\pi/2)$ is a local maximum. Hence, there is a saddle point near $(\pi,\pi/2)$,
or equivalently near $(0,\pi/2)$, $(\pi/2,0)$.
The saddle point gives a very large contribution to the density of states, 
 $N_{\pm}(\epsilon)=\int \left[A_{\pm}(\epsilon,{\bf k})+
B_{\pm}(\epsilon,{\bf k})\right]d^{2}{\bf k}/(2\pi)^{2}$.
Due to the saddle point the electron response function is very strongly peaked
at ${\bf q}\approx (\pi/2,0)$, $\omega \approx 100$meV.
Interestingly, these parameters are very close to where
the anomaly in the breathing phonon mode is observed.~\cite{phonon}
Although further investigation is necessary to clarify this point, we anticipate that the saddle point plays a crucial role in the anomaly.

The quasiparticle residue $Z_{d \bf k\pm}$ is shown in Fig.~\ref{fig:Gd_figure2}(b)  along 
the same line in BZ as dispersion.
Along the nodal direction, the quasiparticle residue $Z_{d\bf k\pm}$ is maximum at the 
bottom of the pocket, and decreases smoothly as moving away from the pocket. 
This smooth dependence is very similar to the residue in undoped parent 
compound.~\cite{Sushkov97} Since the pocket is rather small, the residue inside the pocket 
can be well approximated by a constant $Z_{d\bf k\pm}\approx 0.34$. The bottom of the pocket is well 
fitted by a parabolic band $\epsilon_{\bf k\pm}\approx \beta_{1}k_{1}^{2}/2+\beta_{2}k_{2}^{2}/2$, 
where $\beta_{1}$ and $\beta_{2}$ represent the inverse effective mass along and orthogonal 
to the nodal direction, respectively. 
At the highest doping examined, $x=0.059$, we obtain $\beta_{1}\approx 3.4$, 
$\beta_{2}\approx 1.8$, which yields the anisotropy 
$\beta_{1}/\beta_{2}\approx 1.9$.
The value of the effective mass  $m^{\ast}=1/\sqrt{\beta_{1}\beta_{2}}$ converted to
conventional units is $m^{\ast}\approx 1.6m_{e}$.
This value is consistent with the mass measured in 
MQO.~\cite{Doiron-Leyraud07,LeBoeuf07,Yelland08,Jaudet08,Sebastian08,Audouard09,Sebastian09,Sebastian09_2,Singleton09}

\begin{figure}[ht]
\vspace{-5pt}
\begin{center}
\includegraphics[clip=true,width=0.95\columnwidth]{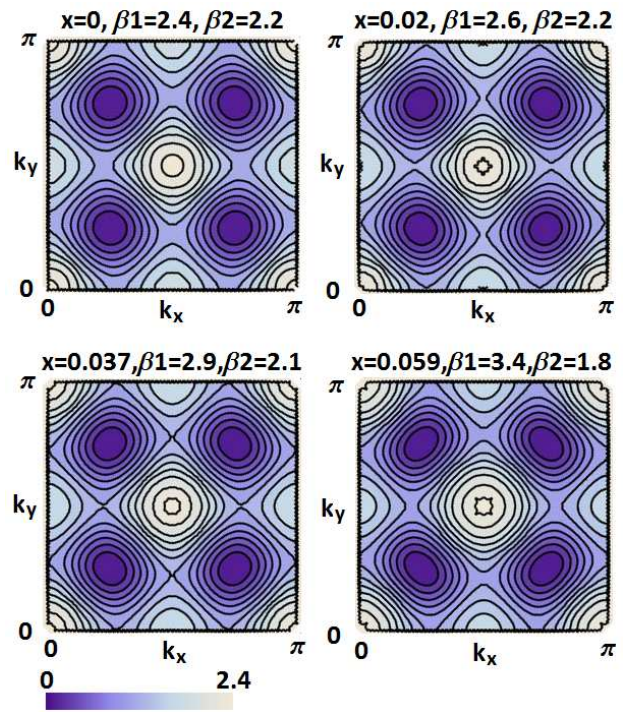}
\caption{(color online) Maps of the parity average dispersion 
$\epsilon_{\bf k}=(\epsilon_{\bf k,+}+\epsilon_{\bf k,-})/2$ at different doping levels. 
Effective mass along ($\beta_{1}$) and orthogonal ($\beta_{2}$) to the nodal direction
is calculated by fitting the bottom of the pocket with 
$\epsilon_{\bf k}\approx \beta_{1}k_{1}^{2}/2+\beta_{2}k_{2}^{2}/2$. 
The ellipticity of hole pockets, represented by $\sqrt{\beta_{1}/\beta_{2}}$, increases 
with doping indicating a deviation from the rigid band  approximation.
}
\label{fig:compare_undoped_doped}
\end{center}
\end{figure}
Maps of the parity average dispersion, $\epsilon_{\bf k}=(\epsilon_{\bf k,+}+\epsilon_{\bf k,-})/2$,
are shown in Fig.~\ref{fig:compare_undoped_doped} for four values of doping.
The total bandwidth $\approx 2.4J\approx 0.3$eV remains roughly the same independent of  doping.
This bandwidth is very close to the value indicated by ARPES in undoped single layer 
compound.~\cite{Wells95,LaRosa97}
By comparing different doping levels in Fig.~\ref{fig:compare_undoped_doped}, it is also clear 
that the ellipticity of the pocket increases with doping. More precisely,
 $\beta_1$ is increasing while $\beta_2$ is decreasing at higher doping.
Enhancement of ellipticity with doping  is consistent with a  previous analytical 
calculation.~\cite{Kotov04}
The doping dependent ellipticity implies that the rigid band  approximation is
strictly speaking not valid. However, the change of ellipticity,
although being significant,  is not dramatic, so the rigid band 
approximation is fairly reasonable.
Moreover, the average effective mass $m^{\ast}=1/\sqrt{\beta_{1}\beta_{2}}$ is 
roughly doping independent.

The present approach certainly supports the small Fermi surface
Luttinger's theorem. The doping calculated according to Eq.~(\ref{doping})
coincides with the area of the small Fermi surface.
It is worth noting that numerically this coincidence is quite nontrivial,
the negative energy incoherent part of the Green's function is absolutely
significant for this.

\section{Spin-charge recombination and ARPES spectral function}

\begin{figure}[ht]
\begin{center}
\includegraphics[clip=true,width=0.95\columnwidth]{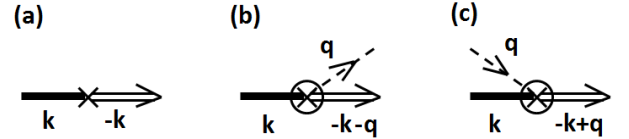}
\caption{Vertices involved in the photoemission process, as defined in Eq.~(\ref{abc_vertices})
}
\label{fig:Gc_vertices}
\end{center}
\end{figure}

In the SCBA approach spin and charge are separated, there are
nonitinerant spins and there are itinerant spinless holes.
The hole has a pseudospin indicating a chequerboard magnetic sublattice, but 
the pseudospin is different from usual spin.
In ARPES process spin and charge recombine to physical electrons.
To calculate the recombination amplitude we
follow the approach of Ref.~\onlinecite{Sushkov97} modifying the approach
to the bilayer case and to  finite doping.
ARPES measures electrons with true spin, regardless which sublattice 
the electron comes from. The annihilation operator of an electron from $m$-th plane ($m=1,2$)
is
\begin{eqnarray}
c_{{\bf k},m,\sigma}=\sqrt{\frac{2}{N}}\sum_{l\in\left\{i,j\right\}}c_{l,m,\sigma}e^{-i{\bf k\cdot r}_{l}}\;,
\label{electron_ft}
\end{eqnarray}
where $\sigma$ is the true spin.
Notice that the definition (\ref{electron_ft}) is properly normalized, because
\begin{eqnarray}
&&\langle 0|c_{{\bf k},1,\uparrow}^{\dag}c_{{\bf k},1,\uparrow}|0\rangle=\frac{2}{N}\langle 0|\sum_{l\in\left\{i,j\right\}}c_{l,1,\uparrow}^{\dag}c_{l,1,\uparrow}|0\rangle
\nonumber \\
&&\;\;\;=\frac{2}{N}\langle 0|\sum_{l\in\left\{i,j\right\}}\left(\frac{1}{2}+S_{l}^{1z}\right)|0\rangle=1\;.
\end{eqnarray}
To establish the connection with hole operators in Eq.~(\ref{hole_parity_basis}), one needs to rotate 
electron operators into the fixed parity basis
\begin{eqnarray}
c_{{\bf k},\pm,\sigma}=\frac{1}{\sqrt{2}}\left(c_{{\bf k},1,\sigma}\pm c_{{\bf k},2,\sigma}\right)
\label{electron_parity_basis}
\end{eqnarray}
The connection between electron and hole Green's function is then associated with the 
following vertices~\cite{Sushkov97,Brink98}
\begin{eqnarray}
a_{\bf k}&=&\langle 0|d_{{\bf -k},+,\uparrow}c_{{\bf k},+,\downarrow}|0\rangle
\nonumber \\
&=&\langle 0|d_{{\bf -k},-,\uparrow}c_{{\bf k},-,\downarrow}|0\rangle
=\sqrt{1/2+m}\;,
\nonumber \\
b_{{\bf k,q},+}&=&\langle 0|\beta_{\bf q,+}d_{{\bf -k-q},+,\downarrow}c_{{\bf k},+,\downarrow}|0\rangle
\nonumber \\
&=&\langle 0|\beta_{\bf q,+}d_{{\bf -k-q},-,\downarrow}c_{{\bf k},-,\downarrow}|0\rangle
=\sqrt{\frac{1}{N}}v_{\bf q,+}
\nonumber \\
b_{{\bf k,q},-}&=&\langle 0|\beta_{\bf q,-}d_{{\bf -k-q},+,\downarrow}c_{{\bf k},-,\downarrow}|0\rangle
\nonumber \\
&=&\langle 0|\beta_{\bf q,-}d_{{\bf -k-q},-,\downarrow}c_{{\bf k},+,\downarrow}|0\rangle
=\sqrt{\frac{1}{N}}v_{\bf q,-}
\nonumber \\
c_{{\bf k,q},+}&=&\langle 0|d_{{\bf -k+q},+,\downarrow}c_{{\bf k},+,\downarrow}\alpha_{\bf q,+}^{\dag}|0\rangle
\nonumber \\
&=&\langle 0|d_{{\bf -k+q},-,\downarrow}c_{{\bf k},-,\downarrow}\alpha_{\bf q,+}^{\dag}|0\rangle
=\sqrt{\frac{1}{N}}u_{\bf q,+}
\nonumber \\
c_{{\bf k,q},-}&=&\langle 0|d_{{\bf -k+q},+,\downarrow}c_{{\bf k},-,\downarrow}\alpha_{\bf q,-}^{\dag}|0\rangle
\nonumber \\
&=&\langle 0|d_{{\bf -k+q},-,\downarrow}c_{{\bf k},+,\downarrow}\alpha_{\bf q,-}^{\dag}|0\rangle
=\sqrt{\frac{1}{N}}u_{\bf q,-}
\label{abc_vertices}
\end{eqnarray}
Here $|0\rangle$ is the ground state of the doped system.
The vertex $a_{\bf k}$ describes the process of instant creation of an electron with
momentum ${\bf k}$ and a hole with momentum $-{\bf k}$, $|0\rangle \to
c_{{\bf k},+,\downarrow}^{\dag}d_{{\bf -k},+,\uparrow}^{\dag}|0\rangle$,
as it is shown in  Fig.~\ref{fig:Gc_vertices}(a).
In the figure electron is shown by the bold solid line  moving from the left, and the hole is 
shown by the double line.
The direction of the electron line in Fig.~\ref{fig:Gc_vertices}
strictly speaking is not correct, as the electron has to appear in the final state only.
Nevertheless, in figures we always show electron in the initial state just
for a convenient graphical presentation.
This way of presentation does not cause any problems since we always have only one electron.
The vertex  $b_{{\bf k,q},\gamma}$
describes the process of instant creation of an electron with
momentum ${\bf k}$, a hole with momentum $-{\bf k}-{\bf q}$ and a magnon
with momentum ${\bf q}$,
as shown in Fig.~\ref{fig:Gc_vertices}(b). We define the parity index $\gamma$ to be parity of the 
magnon. The vertex  $c_{{\bf k,q},\gamma}$ describes the process of annihilation of a magnon
with momentum ${\bf q}$ and instant creation of an electron with
momentum ${\bf k}$ and a hole with momentum $-{\bf k}+{\bf q}$,
as shown in Fig.~\ref{fig:Gc_vertices}(c).
Derivation of the vertices (\ref{abc_vertices}) is presented
in Appendix A. 
The analysis of ARPES in undoped parent compound at zero temperature requires only $a_{\bf k}$ and $b_{{\bf k,q}}$ vertices,~\cite{Sushkov97}
whereas the undoped compound at nonzero temperatures requires additional vertex $c_{\bf k,q}$,~\cite{Brink98} because there are thermally excited magnons in the initial state.
Presently we consider a doped compound at zero temperature, so there are no thermally excited magnetic fluctuations, but there are additional magnetic fluctuations due to doping. Thus all three vertices are
involved in the  present calculation.

\begin{figure}[ht]
\begin{center}
\includegraphics[clip=true,width=0.95\columnwidth]{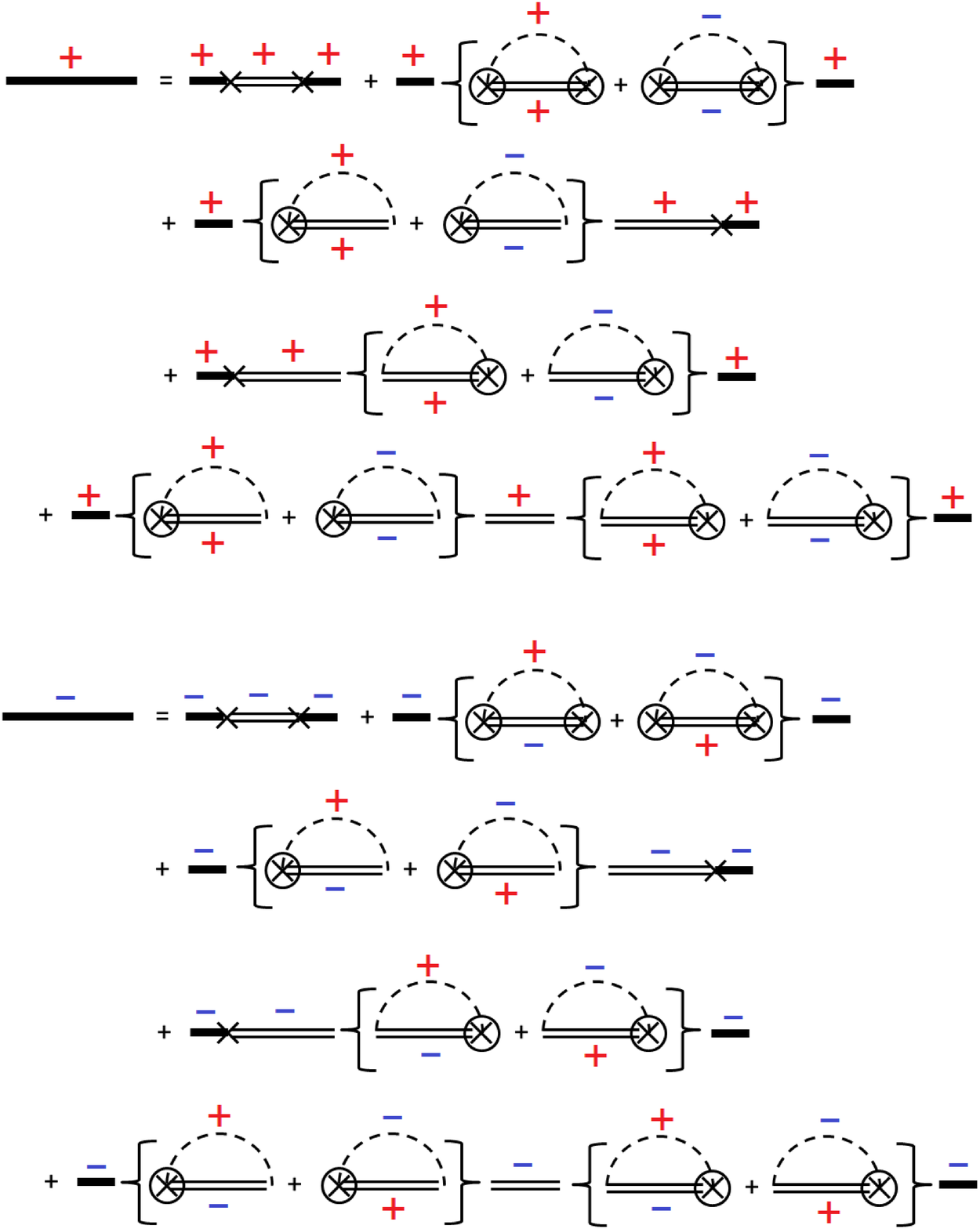}
\caption{(color online) Dyson's equations relating $G_{c\pm}$ and  $G_{d\pm}$.
The thick line represents $G_{c\pm}$, and the double line represents $G_{d\pm}$. }
\label{fig:Gc_Feynman_diagrams}
\end{center}
\end{figure}

The electron Green's function is defined in the standard way.
\begin{eqnarray}
G_{c\sigma\pm}(\epsilon,{\bf k})=-i\int_{-\infty}^{\infty}\langle Tc_{{\bf k},\pm,\sigma}^{\dag}(t)
c_{{\bf k},\pm,\sigma}(0)\rangle e^{i\epsilon t}dt
\label{el_GF}
\end{eqnarray}
Here $\sigma$ is the true spin index.
Notice that the two spins are degenerate, so below we omit the spin index 
in $G_{c\pm}$. 
Dyson equations relating $G_c$ and already calculated $G_d$ are  similar to that
derived in Refs.~\onlinecite{Sushkov97,Brink98}.
The equations are shown graphically in Fig.~\ref{fig:Gc_Feynman_diagrams}
and presented below in analytical form
\begin{eqnarray}
G_{c+}(\epsilon,{\bf k})&=&a_{\bf k}^{2}G_{d+}(\epsilon,{\bf -k})+\Sigma_{++}^{(1)}(\epsilon,{\bf k})+\Sigma_{--}^{(1)}(\epsilon,{\bf k})
\nonumber \\
&+&2a_{\bf k}G_{d+}(\epsilon,{\bf -k})\left[\Sigma_{++}^{(2)}(\epsilon,{\bf k})+\Sigma_{--}^{(2)}(\epsilon,{\bf k})\right]
\nonumber \\
&+&G_{d+}(\epsilon,{\bf -k})\left[\Sigma_{++}^{(2)}(\epsilon,{\bf k})+
\Sigma_{--}^{(2)}(\epsilon,{\bf k})\right]^{2}
\nonumber \\
G_{c-}(\epsilon,{\bf k})&=&a_{\bf k}^{2}G_{d-}(\epsilon,{\bf -k})+\Sigma_{+-}^{(1)}(\epsilon,{\bf k})+\Sigma_{-+}^{(1)}(\epsilon,{\bf k})
\nonumber \\
&+&2a_{\bf k}G_{d-}(\epsilon,{\bf -k})\left[\Sigma_{+-}^{(2)}(\epsilon,{\bf k})+\Sigma_{-+}^{(2)}(\epsilon,{\bf k})\right]
\nonumber \\
&+&G_{d-}(\epsilon,{\bf -k})\left[\Sigma_{+-}^{(2)}(\epsilon,{\bf k})+\Sigma_{-+}^{(2)}(\epsilon,{\bf k})\right]^{2}
\label{Gc_SCBA}
\end{eqnarray}
We emphasize that the conventional self-energy $\Sigma_{\gamma\delta}$ appears only in Dyson's equation 
(\ref{Gd_SCBA}) for $G_d$. Equations (\ref{Gc_SCBA}) contain different
kinds of  self-energy. The self-energy $\Sigma^{(1)}_{\gamma\delta}$ has two
vertices $b_{{\bf k,q},\gamma}$ or $c_{{\bf k,q},\gamma}$ denoted by the crossed circle.
The self-energy $\Sigma^{(2)}_{\gamma\delta}$ has only one
vertex $b_{{\bf k,q},\gamma}$ or $c_{{\bf k,q},\gamma}$ denoted by the crossed circle
and one vertex $g_{\bf k,q}$ (see Eq.~(\ref{hole_magnon_vertices}))
shown by a simple attachment of the dashed line (magnon)
to the double line (hole).
The subscript $\{\gamma\delta\}$ in the self-energies shows parities
of the intermediate magnon and hole, respectively, as displayed in 
Fig.~\ref{fig:Gc_Feynman_diagrams}.
Expressions for $\Sigma_{\gamma\delta}^{(1)}$ and $\Sigma_{\gamma\delta}^{(2)}$ are presented in Appendix B. 
\begin{figure}[ht]
\begin{center}
\includegraphics[clip=true,width=0.95\columnwidth]{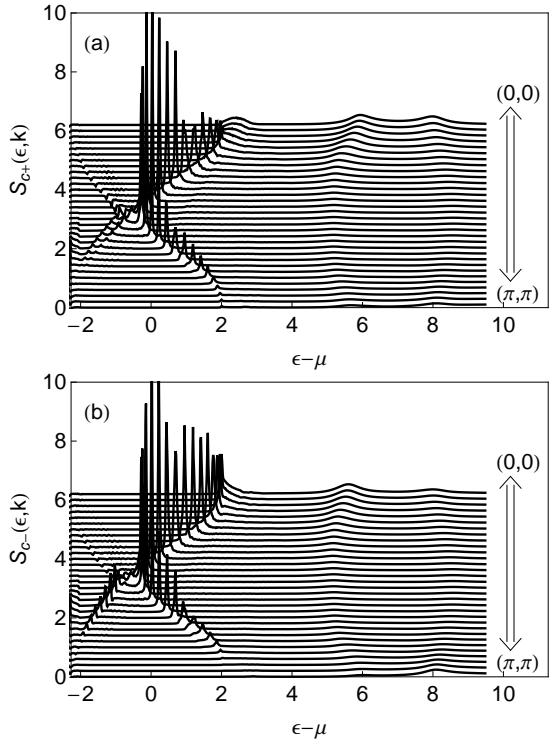}
\caption{Spectral functions $S_{c+}$ (a) and $S_{c-}$ (b) for different values
of ${\bf k}$ along the nodal direction from $(0,0)$ to $(\pi,\pi)$. 
There is an offset 0.2 between subsequent values of ${\bf k}$.
The doping level is $x=0.059$.
}
\label{fig:Gc_figure}
\end{center}
\end{figure}

It is worth noting that meaning of Dyson's equations, Eq.~(\ref{Gc_SCBA}) is different from that of Eq.~(\ref{Gd_SCBA}).
Eq.~(\ref{Gd_SCBA}) is a real dynamic equation for $G_{d\pm}$, which requires an iterative solution.
On the other hand, Eq.~(\ref{Gc_SCBA}) is just a relation between $G_{c\pm}$ 
and $G_{d\pm}$ that describes the spin-charge recombination in the 
photoemission process.
Once $G_{d\pm}$ is known, we calculate the right hand side in Eq.~(\ref{Gc_SCBA}) and 
$G_{c\pm}$ is determined.
The Greens's function $G_{c\pm}$ has the spectral representation similar to
Eq.~(\ref{Gd_spectral_representation}) and a spectral function similar to Eq.~(\ref{sd}).
The spectral functions $S_{c\pm}$ for $x=0.059$ are displayed in 
Fig.~\ref{fig:Gc_figure}(a) and Fig.~\ref{fig:Gc_figure}(b).
The position of the quasiparticle peak is exactly the same as that in $G_{d\pm}$, as one expects from general considerations and it is also evident from Eqs.~(\ref{Gc_SCBA}).
However, the quasiparticle residue of $G_{c\pm}$, denoted by $Z_{c\pm}$, is very different from 
that of $G_{d\pm}$. Since $G_{c\pm}$ represents electrons of either sublattice, there is no a
Bloch theorem to require $Z_{c\pm}$ to be symmetric with respect to MBZ boundary.

\begin{figure}[ht]
\begin{center}
\includegraphics[clip=true,width=0.9\columnwidth]{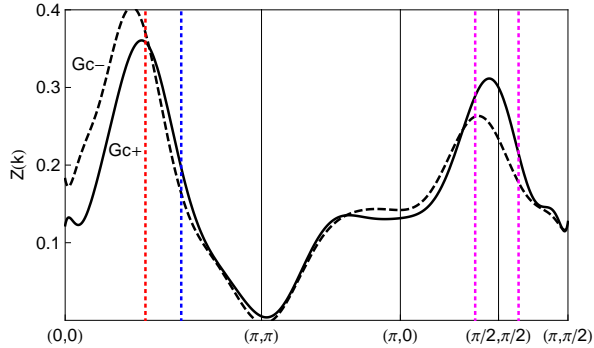}
\caption{(color online) The quasiparticle residue $S_{c}$ along the same line as in 
Fig.~\ref{fig:Gd_figure2}. Doping level is $x=0.059$.
Vertical dashed lines show Fermi points. }
\label{fig:Gc_figure2}
\end{center}
\end{figure}

In fact, our calculation shows a very asymmetric $Z_{c\pm}$, as shown in 
Fig.~\ref{fig:Gc_figure2} along a certain line in BZ.
Along the nodal direction, 
$Z_{c\pm}$ first increases from ${\bf k}=(0,0)$ to the inner edge(red dotted 
vertical line) of the Fermi pocket. After that $Z_{c\pm}$ drops 
abruptly as ${\bf k}$ goes across the pocket inside the Fermi surface
(although this part of the Green's function is not measurable by ARPES).
A less steep drop of $Z_{c\pm}$ takes place
from the outer edge (blue dotted vertical line) of the pocket to the point
${\bf k}=(\pi,\pi)$. The difference between $Z_{c\pm}$ on the inner side of 
the pocket and the outer side is very significant.

\begin{figure}[ht]
\begin{center}
\includegraphics[clip=true,width=0.95\columnwidth]{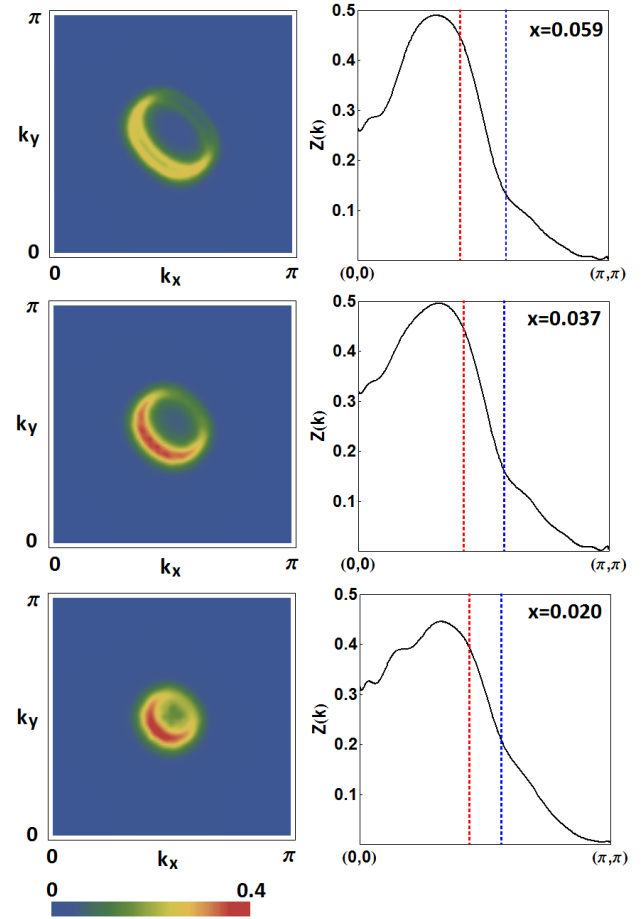}
\caption{(color online) Right column: The parity average residue 
$\left[Z_{c+}^{H}({\bf k})+Z_{c-}^{H}({\bf k})\right]/2$ of the electron Green's
function along the nodal direction.
Vertical dashed lines show Fermi points.
The ARPES intensity is proportional to the residue outside the Fermi surface.
Left column: colour maps of the ARPES intensity at the Fermi surface.
The plots are presented for three values of doping, $x=0.02, 0.037, 0.059$.
}
\label{fig:Gc_Hubbard_figure}
\end{center}
\end{figure}

The above analysis is performed within the double layer 
$t-t^{\prime}-t^{\prime\prime}-J$ model. 
The $t-t^{\prime}-t^{\prime\prime}-J$  model can originate from the single 
band Hubbard model or
from a multi-band Hubbard model. The origin of the
$t-t^{\prime}-t^{\prime\prime}-J$ model is not important for the dynamic
equation (\ref{Gd_SCBA}).
However, the origin is important for the spin-charge recombination
amplitude given by Eq.~(\ref{Gc_SCBA}). Depending on the original model,
there is an additional significant 
contribution to the asymmetry of the electron spectral function between 
inside and outside of MBZ. Analysis performed in Ref.~\onlinecite{Sushkov97}
shows that in the case of the single 
band Hubbard model the vertices in Eq.~(\ref{abc_vertices}) should be modified 
by
\begin{eqnarray}
&&a_{\bf k}\rightarrow a_{\bf k}\left(1+\frac{J}{t}\gamma_{\bf k}\right)\nonumber \\
&&b_{{\bf k,q},\pm}\rightarrow b_{{\bf k,q},\pm}\left(1+\frac{J}{t}\gamma_{\bf k}\right)\nonumber \\
&&c_{{\bf k,q},\pm}\rightarrow c_{{\bf k,q},\pm}\left(1+\frac{J}{t}\gamma_{\bf k}\right)
\end{eqnarray}
which effectively modify quasiparticle residue $Z_{c\pm}$ by
\begin{equation}
Z_{c\pm}\to Z_{c\pm}^H=Z_{c\pm}\left(1+\frac{J}{t}\gamma_{\bf k}\right)\;.
\end{equation}
Following Ref.~\onlinecite{Sushkov97} we call this correction the Hubbard model correction.
Plots of the parity-averaged quasiparticle residue, 
$\left[Z_{c+}^{H}({\bf k})+Z_{c-}^{H}({\bf k})\right]/2$, along the nodal direction are presented in Fig.~\ref{fig:Gc_Hubbard_figure} for three different values of doping.
In the same figure, we also present colour maps of the following function
\begin{eqnarray}
\overline{Z}_{c}^{H}({\bf k})=\frac{\delta^{2}}{2}\left\{\frac{Z_{c+}^{H}(\bf k)}{(\epsilon_{\bf k,+}-\mu)^{2}+\delta^{2}}+\frac{Z_{c-}^{H}(\bf k)}{(\epsilon_{\bf k,-}-\mu)^{2}+\delta^{2}}\right\}\;
\label{ARPES_ZcH}
\end{eqnarray}
This formula gives a way to image the quasiparticle residue at the Fermi surface. The broadening $\delta$ can be taken as a result of disorder and/or experimental resolution, or purely for the sake of imaging. We choose $\delta =0.1$, which corresponds
to the halfwidth $\Gamma =2\delta=26$meV.

From the right column of Fig.~\ref{fig:Gc_Hubbard_figure} we observe that 
at the 
largest doping examined, $x=0.059$, the ARPES intensity in the inner side
(red dotted line) of the pocket is about $4$ times larger than that in 
the outer side (blue dotted line). Comparing plots in different doping 
levels, it is also clear that this asymmetry grows with doping.
 This asymmetry is also reflected in the left column of 
Fig.~\ref{fig:Gc_Hubbard_figure}, where maps of 
$\overline{Z}_{c}^{H}({\bf k})$ at corresponding doping levels is presented. 
One clearly sees that the intensity of $\overline{Z}_{c}^{H}({\bf k})$ displays 
an arc shape along the Fermi surface.

\section{conclusions}

At doping below $5.5\%-6\%$ the bilayer cuprate YBa$_{2}$Cu$_{3}$O$_{6+y}$
is a collinear antiferromagnet. The doping-independent staggered 
magnetization at zero temperature is about $0.6\mu_B$.
This is the maximum value of magnetization allowed by quantum fluctuations of 
localized spins. 
These experimental observations create a unique opportunity for theory to perform a 
controlled calculation of the electron spectral function at doping $x < 0.06$.
In the present work we perform such a calculation within the framework of the extended 
$t-J$ model with account of the Hubbard model corrections.
The calculation employs the self-consistent Born approximation (SCBA) that is parametrically
justified because of the long range AF order with maximum possible staggered magnetization.
To perform the work we have developed/extended the SCBA to the finite doping 
case and to the bilayer system.
The calculation clearly demonstrates that the Fermi surface consists of small hole pockets
centered at $(\pm\pi/2,\pm\pi/2)$.
This conclusion itself is a trivial one since we deal with the system with long range AF
order with maximum possible staggered magnetization. The small pocket is a direct
consequence of Bloch theorem.
What is nontrivial is that we quantify the asymmetry of the ARPES spectral function that is
highly anisotropic at the Fermi surface. In particular at doping about $5\%-6\%$ 
the ARPES intensity in the inner side of the pocket is about $4$ times larger 
than that in
the outer  side.  
Overall the picture resembles Fermi arcs observed in ARPES.

Our analysis shows that the hole band is not quite rigid under doping.
In particular the ellipticity of the pocket increases with doping.
However, the effect of changing ellipticity while being 
significant still is not dramatic, so the rigid band approximation is not that bad.
The hole effective mass averaged over the Fermi surface is practically doping independent,
$m^{\ast}\approx 1.6m_{e}$.

Our calculation demonstrates that due to a saddle point in the hole dispersion
the electronic response of the system is peaked at ${\bf q}\approx (\pi/2,0)$, 
$\omega \approx 100$meV. These parameters are very close to those where
the anomaly is observed in the breathing phonon mode.

We also found that the antiferromagnetic correlations practically destroy the bilayer
bonding-antibonding splitting. More precisely the correlations suppress the splitting
by one order of magnitude. If without account of the correlations the splitting is  about 200meV  then at the doping 6\% the value is reduced down to 20meV.\\

\section{Acknowledgments}

We thank P. Horsch, G. Khaliullin, O. Jepsen, A. I. Milstein, and A. Avella for 
stimulating discussions. 
A significant part of this work was done during  stay of W.C. and O.P.S. at the
Max Planck Institute for Solid State Research, Stuttgart,
and stay of O.P.S. at the Yukawa Institute, Kyoto.
W.C. and O.P.S. are very grateful to colleagues
for hospitality and for stimulating atmosphere.

The work was supported by ARC grant DP0881336.
Computations for this project were performed using the  National Computational 
Infrastructure, under project u66. 

This work was also supported 
by the Humboldt Foundation and by
the Japan Society for Promotion of Science.

\appendix

\section{calculation of hole-magnon vertices}
The hole-magnon vertex  due to $H_{t,t_{\perp}}$ in (\ref{double_layer_H})
contains two parts: the first part comes from in-plane nearest-neighbor hopping
$H_{t}$.  For example, the contribution  of the 1st plane hopping in the case
when all parities are positive is
\begin{widetext}
\begin{eqnarray}
&&\langle 0|\alpha_{{\bf q},+}d_{{\bf k},+,\uparrow}\left|H_{t}^{(1)}\right|d_{{\bf k+q},+,\downarrow}^{\dag}|0\rangle
\nonumber \\
&=&\frac{1}{N(1/2+m)}\langle 0|\alpha_{{\bf q},+}\sum_{j}c_{j,1,\downarrow}^{\dag}e^{-i{\bf k\cdot r}_{j}}
\left(-t\right)\left[c_{i,1,\uparrow}^{\dag}c_{j,1,\uparrow}+c_{i,1,\downarrow}^{\dag}c_{j,1,\downarrow}
\right]\sum_{i}c_{i,1,\uparrow}e^{i{\bf (k+q)\cdot r}_{i}}|0\rangle
\nonumber \\
&=&\frac{t}{N(1/2+m)}\langle 0|\alpha_{{\bf q},+}\left\{\sum_{ij}e^{-i{\bf k\cdot r}_{j}+i{\bf (k+q)\cdot r}_{i}}\left[S_{j,1}^{-}\left(\frac{1}{2}+S_{i,1}^{z}\right)+\left(\frac{1}{2}-S_{j,1}^{z}\right)S_{i,1}^{-}\right]\right\}|0\rangle
\nonumber \\
&=&\frac{t}{N}\langle 0|\alpha_{{\bf q},+}\left\{\sum_{j\xi}e^{i{\bf (k+q)\cdot\xi}}e^{i{\bf q\cdot r}_{j}}b_{j,1}+\sum_{i\xi}e^{i{\bf k\cdot\xi}}e^{i{\bf q\cdot r}_{i}}a_{i,1}^{\dag}\right\}|0\rangle=2t\sqrt{\frac{1}{N}}\left(\gamma_{\bf k}u_{{\bf q},+}+\gamma_{\bf k+q}v_{{\bf q},+}\right)\;,
\end{eqnarray}
where we denote ${\bf r}_{i}={\bf r}_{j}+\xi$, and use the mean field decomposition 
$ c_{j,1,\downarrow}^{\dag}c_{i,1,\uparrow}^{\dag}c_{j,1,\uparrow}c_{i,1,\uparrow}\to- c_{j,1,\downarrow}^{\dag}c_{j,1,\uparrow}
\langle c_{i,1,\uparrow}^{\dag}c_{i,1,\uparrow}\rangle=S_{j,1}^{-}(1/2+m)$. 
The second plane hopping $H_{t}^{(2)}$ gives an equal contribution and altogether this results in the
first term in $g_{{\bf k,q},\gamma\delta}$ in Eq.~(\ref{hole_magnon_vertices}).

The contribution from the interlayer hopping is
\begin{eqnarray}
&&\langle 0|\alpha_{{\bf q},+}d_{{\bf k},2,\uparrow}H_{t_{\perp}}d_{{\bf k+q},1,\downarrow}^{\dag}|0\rangle
\nonumber \\
&=&\frac{-2t_{\perp}}{N(1/2+m)}\langle 0|\alpha_{{\bf q},+}\sum_{i}c_{i,2,\downarrow}^{\dag}e^{-i{\bf k\cdot r}_{i}}\left[c_{i,1,\uparrow}^{\dag}c_{i,2,\uparrow}+c_{i,2,\uparrow}^{\dag}c_{i,1,\uparrow}
+c_{i,1,\downarrow}^{\dag}c_{i,2,\downarrow}+c_{i,2,\downarrow}^{\dag}c_{i,1,\downarrow}\right]
\sum_{i}c_{i,1,\uparrow}e^{i({\bf k+q})\cdot{\bf r}_{i}}|0\rangle
\nonumber \\
&=&\frac{-2t_{\perp}}{N(1/2+m)}\langle 0|\alpha_{{\bf q},+}\sum_{i}e^{i{\bf q\cdot r}_{i}}\left[c_{i,2,\downarrow}^{\dag}c_{i,1,\uparrow}^{\dag}c_{i,2,\uparrow}c_{i,1,\uparrow}
+c_{i,2,\downarrow}^{\dag}c_{i,1,\downarrow}^{\dag}c_{i,2,\downarrow}c_{i,1,\uparrow}\right]|0\rangle
\nonumber \\
&=&\frac{2t_{\perp}}{N}\langle 0|\alpha_{{\bf q},+}\sum_{i}e^{i{\bf q\cdot r}_{i}}\left(S_{i,2}^{-}+S_{i,1}^{-}\right)|0\rangle
=t_{\perp}\sqrt{\frac{2}{N}}\langle 0|\alpha_{{\bf q},+}\left(b_{{\bf -q},2}+a_{{\bf q},1}^{\dag}\right)|0\rangle
=t_{\perp}\sqrt{\frac{1}{N}}\left(u_{{\bf q},+}+v_{{\bf q},+}\right)
\end{eqnarray}
Rotating this to the parity basis, one recovers the second term in $g_{{\bf k,q},\gamma\delta}$
in Eq.~(\ref{hole_magnon_vertices}).

The spin-charge recombination vertices (\ref{abc_vertices}) are calculated in a similar way.
The $a$-vertex is the following:
\begin{eqnarray}
a_{\bf k}&=&\langle 0|d_{{\bf -k},+,\uparrow}c_{{\bf k},+,\downarrow}|0\rangle
=\frac{1}{N\sqrt{1/2+m}}\langle 0|\left(\sum_{j}c_{j,1,\downarrow}^{\dag}e^{i{\bf k\cdot r}_{j}}\sum_{l\in\left\{i,j\right\}}c_{l,1,\downarrow}e^{-i{\bf k\cdot r}_{l}}+\sum_{i}c_{i,2,\downarrow}^{\dag}e^{i{\bf k\cdot r}_{i}}\sum_{l\in\left\{i,j\right\}}c_{l,2,\downarrow}e^{-i{\bf k\cdot r}_{l}}\right)|0\rangle
\nonumber \\
&=&\frac{1}{N\sqrt{1/2+m}}\langle 0|\left(\sum_{j}c_{j,1,\downarrow}^{\dag}c_{j,1,\downarrow}
+\sum_{i}c_{i,2,\downarrow}^{\dag}c_{i,2,\downarrow}\right)|0\rangle
=\sqrt{1/2+m}\ .
\end{eqnarray}
The $b$-vertex reads:
\begin{eqnarray}
b_{{\bf k,q},+}&=&\langle 0|\beta_{\bf q,+}d_{{\bf -k-q},+,\downarrow}c_{{\bf k},+,\downarrow}|0\rangle\nonumber\\
&=&\langle 0|\frac{\beta_{\bf q,+}}{N\sqrt{1/2+m}}\left(\sum_{i}c_{i,1,\uparrow}^{\dag}e^{i{\bf (k+q)\cdot r}_{i}}\sum_{l\in\left\{i,j\right\}}c_{l,1,\downarrow}e^{-i{\bf k\cdot r}_{l}}
+\sum_{j}c_{j,2,\uparrow}^{\dag}e^{i{\bf (k+q)\cdot r}_{j}}\sum_{l\in\left\{i,j\right\}}c_{l,2,\downarrow}e^{-i{\bf k\cdot r}_{l}}\right)|0\rangle
\nonumber \\
&=&\langle 0|\frac{\beta_{\bf q,+}}{N\sqrt{1/2+m}}\left(\sum_{i}a_{i,1}e^{i{\bf q\cdot r}_{i}}+\sum_{j}a_{j,2}e^{i{\bf q\cdot r}_{j}}\right)|0\rangle
=\sqrt{\frac{1}{N}}\frac{v_{\bf q,+}}{\sqrt{1/2+m}}\approx\sqrt{\frac{1}{N}}v_{\bf q,+}\;.
\end{eqnarray}
Similarly the $c$-vertex is:
\begin{eqnarray}
c_{{\bf k,q},+}=\langle 0|d_{{\bf -k+q},+,\downarrow}c_{{\bf k},+,\downarrow}\alpha_{\bf q,+}^{\dag}|0\rangle
=\sqrt{\frac{1}{N}}\frac{u_{\bf q,+}}{\sqrt{1/2+m}}\approx\sqrt{\frac{1}{N}}u_{\bf q,+}\;.
\end{eqnarray}

\section{The hole self-energy}
Here we present explicit expressions for each self-energy in the Dyson's equations for $G_{d\pm}$ and $G_{c\pm}$ 
\begin{eqnarray}
\label{hole_magnon_S}
&&\Sigma_{\gamma\delta}(\epsilon,{\bf k})
=\int\frac{d^{2}{\bf q}}{(2\pi)^{2}}g_{\bf k-q,q,\gamma\delta}^{2}
\int_{0}^{\infty}dx\frac{A_{\delta}(x,{\bf k-q})}{\epsilon-\omega_{\bf q,\gamma}-x+i0}
+\int\frac{d^{2}{\bf q}}{(2\pi)^{2}}g_{\bf k,-q,\gamma\left(\gamma\cdot\delta\right)}^{2}\int_{-\infty}^{0}dx\frac{B_{\delta}(x,{\bf k-q})}{\epsilon+\omega_{\bf q,\gamma}-x-i0}\;, \\
&&\Sigma_{\gamma\delta}^{(1)}(\epsilon,{\bf k})
=\int\frac{d^{2}{\bf q}}{(2\pi)^{2}}b_{\bf k,q,\gamma}^{2}
\int_{0}^{\infty}dx\frac{A_{\delta}(x,{\bf k-q})}{\epsilon-\omega_{\bf q,\gamma}-x+i0}
+\int\frac{d^{2}{\bf q}}{(2\pi)^{2}}c_{\bf k,q,\gamma}^{2}\int_{-\infty}^{0}dx\frac{B_{\delta}(x,{\bf k-q})}{\epsilon+\omega_{\bf q,\gamma}-x-i0}\;,
\nonumber \\
&&\Sigma_{\gamma\delta}^{(2)}(\epsilon,{\bf k})
=\int\frac{d^{2}{\bf q}}{(2\pi)^{2}}b_{\bf k,q,\gamma}g_{\bf k-q,q,\gamma\delta}
\int_{0}^{\infty}dx\frac{A_{\delta}(x,{\bf k-q})}{\epsilon-\omega_{\bf q,\gamma}-x+i0}
+\int\frac{d^{2}{\bf q}}{(2\pi)^{2}}c_{\bf k,q,\gamma}g_{\bf k,-q,\gamma\left(\gamma\cdot\delta\right)}\int_{-\infty}^{0}dx\frac{B_{\delta}(x,{\bf k-q})}{\epsilon+\omega_{\bf q,\gamma}-x-i0}\;,
\nonumber
\end{eqnarray}
We remind that $\left\{\gamma,\delta\right\}=\left\{\pm,\pm\right\}$.
Notice that due to violation of cross-leg symmetry, the retarded and advanced part have different vertices, 
as shown in Fig.~\ref{fig:Gd_Feynman_diagrams}, see also Refs.~\onlinecite{Igarashi92,Kyung96}. 
It is worth mentioning that $\Sigma_{\gamma\delta}$ is the true self-energy, while $\Sigma_{\gamma\delta}^{(1)}$ 
and $\Sigma_{\gamma\delta}^{(2)}$ are dimensionless quantities 
describing the spin-charge recombination process. 

Direct evaluation of (\ref{hole_magnon_S}) requires a three dimensional integration, two momenta
and one frequency. This is too expensive computationally. Fortunately, using the Kramers-Kronig
dispersion relation one can effectively remove one integration.
 Take $\Sigma_{+-}$ for example, its imaginary part is
\begin{eqnarray}
&&Im\Sigma_{+-}(\epsilon,{\bf k})
\nonumber \\
&&=\pi\int\frac{d^{2}{\bf q}}{(2\pi)^{2}}\left[
-g_{\bf k-q,q,+-}^{2}\int_{0}^{\infty}dx A_{-}(x,{\bf k-q})\delta(\epsilon-\omega_{\bf q,+}-x)
+g_{\bf k,-q,+-}^{2}\int_{-\infty}^{0}dx B_{-}(x,{\bf k-q})\delta(\epsilon+\omega_{\bf q,+}-x)\right]
\nonumber \\
&&=\pi\int\frac{d^{2}{\bf q}}{(2\pi)^{2}}\left[-g_{\bf k-q,q,+-}^{2}
A_{-}(\epsilon-\omega_{\bf q,+},{\bf k-q})\theta(\epsilon-\omega_{\bf q,+})
+g_{\bf k,-q,+-}^{2} B_{-}(\epsilon+\omega_{\bf q,+},{\bf k-q})\theta(-\epsilon-\omega_{\bf q})\right]
\nonumber \\
&&=Im\Sigma^{A}_{+-}(\epsilon,{\bf k})+Im\Sigma^{B}_{+-}(\epsilon,{\bf k})\;.
\label{ImS}
\end{eqnarray}
Evaluation of $Im\Sigma^{A}$ and $Im\Sigma^{B}$ requires only a two-dimensional
 integration over
momenta. The Kramers-Kronig relation is then applied to find the real part
by the principal value integration
\begin{eqnarray}
Re\Sigma^{A}_{+-}(\epsilon,{\bf k})=-\int\frac{d\epsilon^{\prime}}{\pi}\frac{Im\Sigma^{A}_{+-}(\epsilon^{\prime},{\bf k})}{\epsilon-\epsilon^{\prime}}\;,
\ \ \ \ \ \ \
Re\Sigma^{B}_{+-}(\epsilon,{\bf k})=\int\frac{d\epsilon^{\prime}}{\pi}\frac{Im\Sigma^{B}_{+-}(\epsilon^{\prime},{\bf k})}{\epsilon-\epsilon^{\prime}}\;.
\label{ReS}
\end{eqnarray}

\end{widetext}


\begin{thebibliography}{99}
\bibitem{com1} La$_{2-x}$Sr$_x$CuO$_4$ is a superconductor at
$x> 0.055$, but still it is an Anderson insulator from the
point of view of the single particle physics.
\end{thebibliography}
\end{document}